\documentclass[sigconf]{acmart}

\renewcommand\footnotetextcopyrightpermission[1]{} 
\usepackage{graphicx}
\usepackage{amsmath}
\usepackage{algorithm}
\usepackage[noend]{algpseudocode}
\usepackage{tabularx}
\usepackage{verbatim}
\usepackage{chngpage}

\setcopyright{none}

\acmConference[Working paper]{}
\acmYear{2018}
\copyrightyear{}

\acmArticle{}
\acmPrice{}

\begin{document}
\title{Role action embeddings: scalable representation of network positions}

\author{George Berry}
\orcid{}
\affiliation{%
  \institution{Facebook}
}
\email{berry@fb.com}

\renewcommand{\shortauthors}{G. Berry}

\begin{abstract}
We consider the question of embedding nodes with similar local neighborhoods together in embedding space, commonly referred to as ``role embeddings.'' We propose \textsc{RAE}, an unsupervised framework that learns role embeddings. It combines a within-node loss function and a graph neural network (GNN) architecture to place nodes with similar local neighborhoods close in embedding space. We also propose a faster way of generating negative examples called \textit{neighbor shuffling}, which quickly creates negative examples directly within batches. These techniques can be easily combined with existing GNN methods to create unsupervised role embeddings at scale. We then explore \textit{role action embeddings}, which summarize the non-structural features in a node's neighborhood, leading to better performance on node classification tasks. We find that the model architecture proposed here provides strong performance on both graph and node classification tasks, in some cases competitive with semi-supervised methods.

\end{abstract}

%
%

\maketitle

\section{Introduction}

Recently, work on (structural) role embeddings \cite{donnat_learning_2018,ribeiro_struc2vec:_2017} has returned to fundamental questions about node positions in networks posed in the sociological networks literature \cite{harrison_c._white_social_1976,burt_cohesion_1978}. In contrast to community-based embeddings like DeepWalk/node2vec \cite{perozzi_deepwalk_2014,grover_node2vec:_2016} which represent network neighbors close in embedding space, role embeddings place nodes with similar local structures close in embedding space. For instance, when embedding a social network, a community-based embedding places two friends close in embedding space, while role embedding places two people with similar local networks close in space (regardless of whether they know one another).

Rapid progress on graph neural networks (GNNs) \cite{kipf_semi-supervised_2016,hamilton_representation_2017} offers the possibility of scaling role embeddings to large graphs \cite{ying_graph_2018}. To learn role embeddings with GNNs, we propose an unsupervised \textit{within-node} loss function which places nodes close in embedding space when their local neighborhood structures are similar. This allows GNNs to produce role embeddings similar to analytical methods \cite{donnat_learning_2018}, while remaining inductive and more scalable than comparable role embedding techniques \cite{ribeiro_struc2vec:_2017,henderson_rolx:_2012}. By simply changing the loss function for an unsupervised GNN, we can improve its ability to learn role embeddings.

We also introduce the concept of \emph{role action embeddings}, in contrast to the more familiar \emph{structural role embeddings}. By ``action'', we mean any non-structural features of nodes, such as the words used by a paper in a citation graph. Whereas structural role embeddings \cite{donnat_learning_2018,ribeiro_struc2vec:_2017,henderson_rolx:_2012} place nodes with similar network neighborhoods close in embedding space, role action embeddings propagate node actions along the graph, representing nodes as similar when their local neighborhoods contain similar action profiles filtered through similar structures. In addition to being a meaningful theoretical distinction, this choice has practical significance: it improves performance on node classification tasks. Table \ref{table:overview} lays out where this paper fits in with recent literature on node embeddings.

To implement role embeddings with the loss function proposed, we use a modeling framework which we call \textsc{RAE} (short for ``role action embeddings'')\footnote{Code used in this paper may be found at \url{https://github.com/georgeberry/role-action-embeddings}.}. \textsc{RAE} is based on \textsc{GraphSAGE} \cite{hamilton_representation_2017} with several important distinctions. It overcomes the underfitting problem sometimes observed with unsupervised GNNs \cite{xu_how_2018} while being simpler than the standard \textsc{GraphSAGE} model. \textsc{RAE} achieves strong results on node and graph classification tasks, competitive with semi-supervised methods for the former and outperforming more complex kernel methods for the latter.

\begin{figure*}
 \centering
 \includegraphics[width=0.8\textwidth]{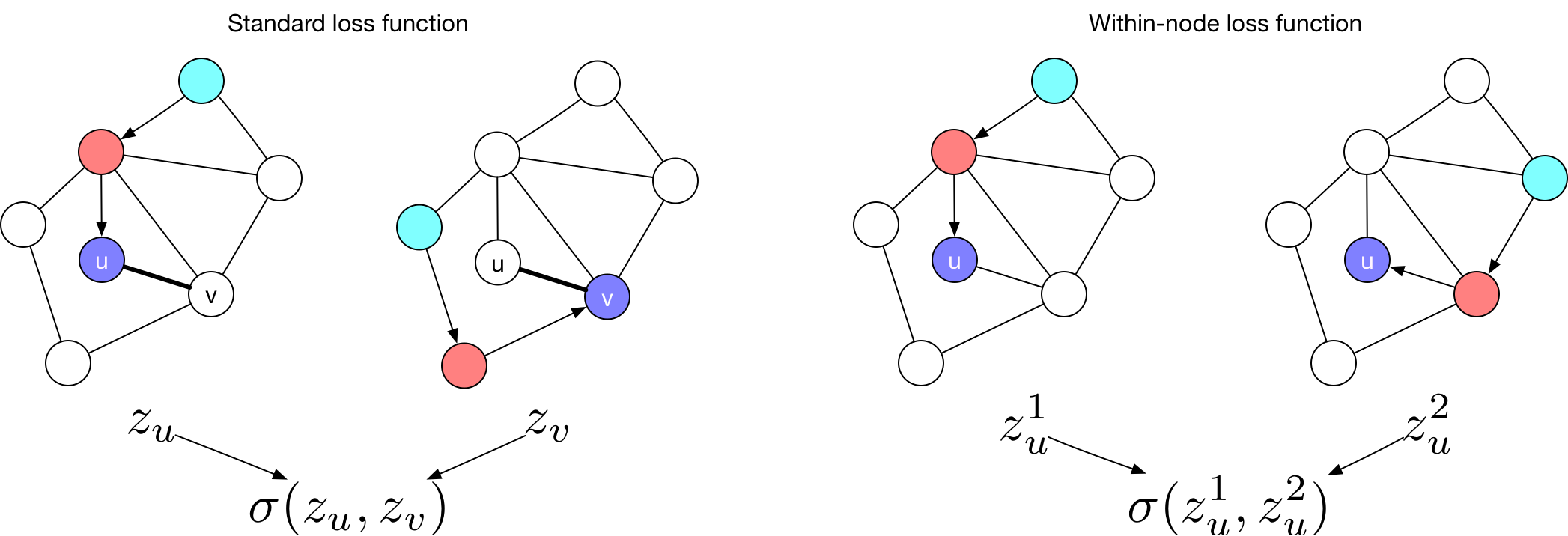}
 \caption{On the left an illustration of a standard unsupervised loss function, similar to \cite{hamilton_representation_2017, grover_node2vec:_2016, perozzi_deepwalk_2014}. We term this $J_{neighbor}$ because it creates positive samples from neighbors in the graph: here, $u$ and $v$ are neighbors so $v$ is a positive example for $u$. On the right is the within-node loss function $J_{within}$ we propose in this paper. It uses as positive examples two samples from $u$'s own neighborhood. Nodes are colored by distance from a focal node, indicating that GNNs iteratively summarize further reaches of a focal node's neighborhood. }
 \label{figure:example}
\end{figure*}

\setlength{\extrarowheight}{.2em}
\begin{table*}
\begin{tabularx}{0.8\textwidth}{lllX}
Goal & Type & Features & Method                                                                                       \\
\hline \hline
Role           & Structural         &  $A, X_s$    & GraphWAVE \cite{donnat_learning_2018}, struc2vec \cite{ribeiro_struc2vec:_2017}, role2vec \cite{ahmed_learning_2018}, RolX \cite{henderson_rolx:_2012}                                        \\
Role           & Action             &   $A, X_a$   & Role action embeddings (\textsc{RAE}, this paper)\\
Community      & Structural         &  $A, X_s$    & DeepWalk \cite{perozzi_deepwalk_2014}, node2vec \cite{grover_node2vec:_2016}, unsupervised GraphSAGE (structural features) \cite{hamilton_representation_2017} \\
Community      & Action             &  $A, X_a$    & unsupervised GraphSAGE (action features) \cite{hamilton_representation_2017}\\
\hline
\end{tabularx}
\caption{Different node embedding types. Role embeddings place nodes with similar local neighborhoods close in embedding space. Community embeddings place nodes which tend to be linked close in embedding space.}\label{table:overview}
\end{table*}

\section{Contributions}
\begin{itemize}
\item We propose a within-node loss function which creates high-quality role embeddings, is compatible with scalable GNN architectures, and requires less graph information than an adjacency-based loss function
\item We propose an unsupervised model architecture which learns quickly and is simpler than many alternatives
\item We show that, for node classification tasks, focusing on action vectors of nodes leads to increased unsupervised performance over the common practice of concatenating action and structural features
\item We introduce neighbor shuffling to quickly create training examples within batches
\item We evaluate this framework (called \textsc{RAE}) in a variety of settings, finding good performance on both node and graph classification tasks
\end{itemize}

\section{Setting and model}
We have a graph $G = (V, E)$ which is treated as undirected. Each node in $G$ has a vector of attributes $X = (X_s, X_a)$, which can be divided into structural features $X_s$ and action features $X_a$. Structural features can include degree, clustering coefficient, and centrality measures. However, we assume only node degrees are readily available. Action features are characteristics of nodes, such as words in documents, chemical properties of molecules, or actions of people.

The goal is to find $d$-dimensional embeddings $z_u \in R^d$ for each node $u \in V$. Importantly, $z_u$ is required to be \textit{inductive}, meaning embeddings of unseen nodes can be obtained.

We build on the \textsc{GraphSAGE} framework \cite{hamilton_representation_2017} for this task. \textsc{GraphSAGE} takes a depth parameter $K$ and has two basic operations: \textsc{combine} and \textsc{aggregate}. At each depth $k \in \{1 \dots K\}$, the representation from the previous layer $h_u^{k-1}$ is updated by sampling neighbors $\text{N}(u)$, and performing \textsc{aggregate} on neighbor features $\{h_v^{k-1} \; \forall v \in \text{N}(u)\}$. Then, this aggregated representation is combined with the existing representation to obtain 
\[
h_u^k = \textsc{combine}(h_u^{k-1}, \textsc{aggregate}(\{h_v^{k-1} \; \forall v \in \text{N}(u)\}).
\]

\noindent
Note that usually $h_u^0 = x_u$.

There are many possible choices for both \textsc{combine} and \textsc{aggregate} \cite{hamilton_representation_2017,xu_how_2018,chen_stochastic_2017}. Common choices for \textsc{combine} are concatenation and mean, and common choices for \textsc{aggregate} are mean, max pool, summation, and LSTM. A fuller discussion of the strengths of different frameworks can be found in \cite{xu_how_2018}. The \textsc{GraphSAGE} framework is in some ways comparable to the Graph Convolutional Network (GCN) framework of \cite{kipf_semi-supervised_2016,chen_fastgcn:_2018}, although it is more scalable \cite{ying_graph_2018}.

We propose a modification of \textsc{GraphSAGE} called \textsc{RAE}. There are several important differences from a standard \textsc{GraphSAGE} model. First, we ignore the \textsc{combine} operation entirely, and always include $u$ whenever the neighbor function $\text{N}(u)$ is used. For clarity, we write $\{u\} \cup \{x_v \: \forall \: v \in \text{N}(u)\}$. In combination with an elementwise mean \textsc{aggregate} function, this has the effect of blending together $u$'s embedding with its neighbors,.

Second, \textsc{RAE} uses a $\tanh$ activation function rather than the standard ReLU. Through experiments, we found dramatically increased unsupervised model performance. Algorithms \ref{algorithm:RAE} and \ref{algorithm:meanagg} lay out the specifics of these choices.

Additionally, we make two practical decisions which improve performance. First, when generating two embeddings $z_u$ and $z_v$ which will be multiplied in a loss function, we generate $z_u$ and $z_v$ from two separate models, similarly to word2vec. Finally, GNNs have the powerful property of providing a distinct embedding at each aggregation level $\{1 \dots K\}$. We specify $K = 2$ below, and use both first and second step embeddings for prediction. This additional information improves model performance and is essentially free, since no additional model training is needed.

\begin{algorithm}
\caption{GNN architecture}\label{algorithm:RAE}
\textbf{Inputs}: Graph $G = (V, E)$; action features $\{x_u^a \; \forall u \in V\}$; depth $K$; batch normalizers $B^k$; mean aggregators $\textsc{Agg}^k$; weight matrices $W^k$; dropout $D$; nonlinearity layer $\sigma^k = D(\text{tanh}(B^k(.)))$ \\
\textbf{Outputs}: $d$-dimensional embeddings $z_u \; \forall u \in V$ \\
\begin{algorithmic}[1]
\Procedure{RAE}{}
\State $h_u^0 \leftarrow x_u^a \: \forall \: u \in V$
\For{$k \in \{1, \dots, K\}$}
\For{$u \in V$}
\State $h_{N(u)}^k \leftarrow \textsc{Agg}^k(u)$
\State $h_u^k \leftarrow \sigma^k(W^k h_{N(u)}^k)$
\EndFor
\EndFor
\State $z_u \leftarrow h_u^K \; \forall u \in V$
\EndProcedure
\end{algorithmic}
\end{algorithm}

\begin{algorithm}
\caption{Mean aggregator}\label{algorithm:meanagg}
\textbf{Inputs}: Focal node $u$; graph $G = (V, E)$; arbitrary features $\{x_u \; \forall u \in V\}$; elementwise mean function $\textsc{Mean}(.)$; batch normalizer $B$; weight matrix $W$; neighborhood sampler $N(u)$; dropout $D$; nonlinearity layer $\sigma = D(\text{tanh}(B(.)))$ \\
\textbf{Outputs}: $d$-dimensional neighbor summary for $u$, $h_{N(u)} \forall u \in V$ \\
\begin{algorithmic}[1]
\Procedure{MeanAggregator}{}
\State $h_{N(u)} \leftarrow \textsc{Mean}(\{u\} \cup \{x_v \: \forall \: v \in \textsc{N}(u)\})$
\State $h_{N(u)} \leftarrow \sigma(W \cdot h_u)$
\State \textbf{return} $h_{N(u)}$
\EndProcedure
\end{algorithmic}
\end{algorithm}

\subsection{Within-node loss function for role embeddings}

Consider two neighboring nodes, $(u, v) \in E$. The unsupervised loss function proposed by \cite{hamilton_representation_2017} seeks to place $u$ and $v$ close in embedding space, by treating $z_v$ as a positive example for $z_u$,

\begin{equation}
- J_{neighbor} = \sigma(z_u^\text{T} z_v) + Q \, \mathbb{E}_{u_n \sim P(u)} [\sigma(-z_u^\text{T} z_{u_n})]
\end{equation}

We refer to this as the ``between node'' or ``neighbor'' loss function. Here, $\sigma$ represents the logsigmoid transformation, $Q$ is the number of negative samples, and $P(u)$ samples a random node not adjacent to $u$. The intuition is close to that from word2vec \cite{mikolov_efficient_2013} and associated methods applied to graphs such as DeepWalk \cite{perozzi_deepwalk_2014} and node2vec \cite{grover_node2vec:_2016}. Essentially, $J_{neighbor}$ treats $u$ as a collection of its neighbors $v$.

An alternate way to think about $u$'s position in embedding space is as a collection of substructures in its own local network neighborhood. Assume we take two samples from $u$'s local neighborhood $u_1$ and $u_2$ to create embeddings $z_{u_1}$ and $z_{u_2}$. We'd like these embeddings to be close in space, while the embedding of a walk from a random node random node $z_w$, $w \neq u$, should be more distant. The intuition here is closer to deep graph kernel techniques \cite{narayanan_graph2vec:_2017,narayanan_subgraph2vec:_2016,yanardag_deep_2015}. Let $R(u)$ sample any $w \neq u$ at random. Then, we can create a within-node loss function
\begin{equation}
- J_{within} = \sigma(z_{u_1}^\text{T} z_{u_2}) + Q \, \mathbb{E}_{w \sim R(u)} [\sigma(-z_{u_1}^\text{T} z_{w})].
\end{equation}

\begin{figure}
 \centering
 \includegraphics[width=0.4 \textwidth]{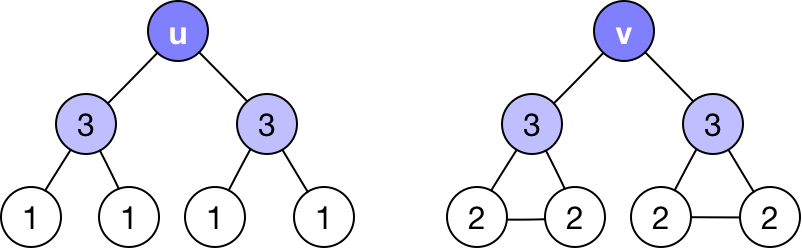}
 \caption{Local neighborhoods of nodes $u$ and $v$. Note that when considering node degree as a structural feature of nodes, the first order neighborhoods (light blue) are identical but the second order neighborhoods (white) are different. When choosing a depth of $K=1$, $J_{within}$ would consider $u$ and $v$ similar, while $J_{neighbor}$ would consider $u$ and $v$ different. This happens because $J_{neighbor}$ embeds nodes together when their neighbors (light blue) have similar $K$-hop neighborhoods. On the other hand, $J_{within}$ looks directly at the $K$-hop neighborhood of the focal nodes.}
 \label{figure:counterexample}
\end{figure}

$J_{within}$ will place nodes close in embedding space when they themselves have similar local neighborhoods. For some graph structures, both $J_{neighbor}$ and $J_{within}$ will lead to similar embeddings, but this is not true in all cases. A simple example of divergence can be seen in Figure \ref{figure:counterexample}, where for depth $K=1$, $J_{neighbor}$ would consider $u$ and $v$ different since two-hop neighbors have different degrees. On the other hand, $J_{within}$ would consider $u$ and $v$ similar since the $K$-step neighborhoods are identical.

This highlights an important difference between $J_{neighbor}$ and $J_{within}$: $J_{within}$ requires relatively less graph information for each training pair since it relies on the $K$ step neighborhood of $u$ rather than the $K+1$ step neighborhood. This offers the possibility of faster training for large graphs, for instance by parallelizing individual neighborhoods. Below, we find that $J_{within}$ produces better role embeddings on ground-truth graphs when measured by silhouette score compared to $J_{neighbor}$.

\subsection{Neighbor shuffling}

We expect embeddings based on two samples from $u$'s neighborhood to naturally be more similar than a sample from $u$ and from a random node $v$. If it is too easy for the model to distinguish samples from $u$ and negative samples from $v$, this could limit embedding quality.

To address this we introduce \textit{neighbor shuffling} to create harder negative examples. This uses the the neighbors of some random node $v$ instead of $u$ to create negative examples. Practically, this means that we choose some $v \neq u$ in line 2 of Algorithm \ref{algorithm:meanagg}. Neighbor shuffling can be easily implemented within batches by permuting the within-batch adjacency list. This approach is similar to the idea of a corruption function in \cite{velickovic_deep_2018} with an important distinction: we shuffle the neighbors rather than the features of $u$, which we found leads to better performance with \textsc{RAE}.

\section{Role embedding performance on exemplar graphs}

\begin{figure*}
 \centering
 \includegraphics[width=\textwidth]{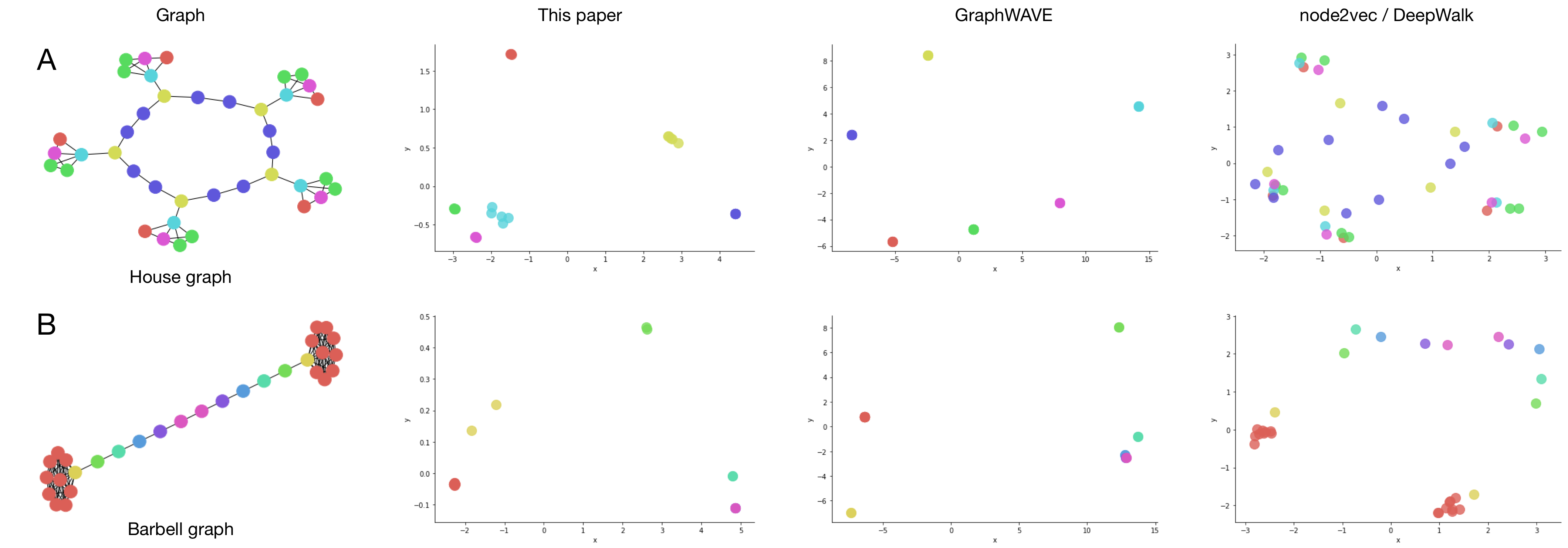}
 \caption{Performance of $J_{within}$ on two graphs with clear ground truth, the house and the barbell. Feature vectors used here are sorted degrees of neighbors. We take GraphWAVE (third column) as ground truth for this task. Note the intuitive difference between role embeddings (columns 2 and 3) and community-based embeddings (column 4).}
 \label{figure:smallgraphs}
\end{figure*}

Past work on structural role embeddings has studied the barbell and house graphs as visual tests of role embedding performance \cite{ribeiro_struc2vec:_2017,donnat_learning_2018}. Figure \ref{figure:smallgraphs} displays our approach in comparison to two alternate methods. We use depth $K = 2$, sorted vectors of neighbor degrees as node features, and train for 100 epochs. We sample 2 neighbors at depth $K=1$ and 4 neighbors at depth $K = 2$.

We compare to GraphWAVE \cite{donnat_learning_2018} and node2vec/DeepWalk \cite{goyal_graph_2017,perozzi_deepwalk_2014}. GraphWAVE is a matrix factorization approach to structural role embeddings based on heat diffusion. Because of GraphWAVE's strong performance for role embeddings, we consider it close to ground truth for this task. node2vec/DeepWalk provides a contrast between role embeddings and community-based embeddings. As expected, the structural role embeddings produced by our method display some variance when compared to GraphWAVE, but they reproduce the pattern of that strong baseline.

\begin{table}
\centering
\begin{tabular}{lll}
& $J_{within}$ & $J_{neighbor}$ \\
\hline \hline
House graph   & $91.2 \pm 2.9\%$ & $89.6 \pm 4.2\%$ \\
Barbell graph & $69.6\pm 8.1\%$ & $69.5 \pm 6.8\%$ 
\end{tabular}
\caption{Silhouette scores and standard errors for ground-truth roles when using identical model setups with different loss functions, from 50 model runs. The differences are statistically significant for the house graph, using a two-sample two-sided t-test: $t = 2.19, df = 98, p < 0.05$.}
\label{table:silhouette}
\end{table}

We compare the embeddings produced by $J_{within}$ and $J_{neighbor}$ in Table \ref{table:silhouette} on otherwise similar models using silhouette scores, finding that $J_{within}$ ranks higher, particularly for the house graph. This makes sense, since $J_{neighbor}$ represents nodes as combinations of their neighbors, leading to somewhat higher variance within roles. Both loss functions produce reasonable role embeddings in combination with structural features.

\section{Experiments}

We now turn to the performance of \textsc{RAE} on two types of tasks: node classification and graph classification.

Intuitively, good representations of a node's neighborhood should allow classifying nodes well in an unsupervised setting. Further, the combination of embeddings for all nodes in a graph should provide distinct graph vectors suitable for graph classification. This treats a graph as a collection of roles. We note that these two tasks put \textsc{RAE} in comparison with two distinct lines of research: the first on node embeddings \cite{grover_node2vec:_2016,perozzi_deepwalk_2014,hamilton_representation_2017} and the second on graph kernels \cite{yanardag_deep_2015,shervashidze_weisfeiler-lehman_2011,niepert_learning_2016,verma_graph_2018}.

\begin{table}
\centering
\begin{tabular}{llll}
& Cora & Citeseer & Pubmed \\
\hline
\hline
Nodes & 2,708 & 3,327 & 19,717 \\
Edges & 5,429 & 4,732 & 44,338 \\
Features & 1,433 & 3,703 & 500 \\
Classes & 7 & 6 & 3 \\
Train/Val/Test & 140/500/1,000 & 120/500/1,000 & 60/500/1,000 \\
\hline
\end{tabular}
\caption{Node classification datasets. The train/validation/test sets splits are set by \cite{kipf_semi-supervised_2016}.}
\label{table:node_datasets}
\end{table}

\begin{table*}[]
\begin{tabular}{lllllll}
Method       & MUTAG & IMDB-B & REDDIT-B & IMDB-M & REDDIT-M5K & REDDIT-M12K \\
\hline \hline
\textbf{Baselines} & & & & & & \\

DGK \cite{yanardag_deep_2015}         & 87.4  & 67.0   & 78.0 & 44.6 & 41.3 &   32.2 \\
\textsc{Patchy-san} \cite{niepert_learning_2016}   & 92.6  & 71.0   & 86.3 & 45.2 & 49.1 & 41.3   \\
GIN \cite{xu_how_2018}   & 90.0  & 75.1   & 92.4 & 52.3 & 57.5 &  -  \\
\hline

\textbf{\textsc{RAE} (this paper)} & & & & & & \\
Untrained & $88.1 \pm 0.9\%$ & $72.7 \pm 0.9\%$ & $81.6 \pm 1.4\%$ & $50.6 \pm 0.5\%$ & $54.3 \pm 1.4\%$ & $43.5 \pm 1.8\%$ \\
\hline

\end{tabular}
\centering
\caption{Results for graph classification tasks with standard deviations from 20 runs. Note that GIN \cite{xu_how_2018} provides an upper bound for expressiveness with a GNN framework.}
\label{table:graph_results}
\end{table*}

\subsection{Model setup}

\begin{table*}[h]
\begin{adjustwidth}{-2in}{-2in} 
\begin{center}
\begin{tabular}{lllllll}
Information   & Method              & d & K & Cora               & Citeseer                      & Pubmed                        \\
\hline \hline
\textbf{Unsupervised baselines} & & & & & \\
$X_a$         & L2 regression     & $|X_a|$ & 0 & 58.9               & 60.4                          & 72.9                          \\
$A $          & DeepWalk \cite{kipf_semi-supervised_2016}  & 128 & 5 & 67.2               & 43.2                          & 65.3                          \\
$A, X_a$      & DeepWalk + $X_a$ \cite{hamilton_representation_2017}  & 128 + $|X_a|$ & 5 & 70.7    & 51.4                          & 74.3                          \\
$A, X_a$      & DGI \cite{velickovic_deep_2018} & 512 (256 on Pubmed) & 3  & \textbf{82.3}            & \textbf{71.8}              & 76.8                          \\
$A, X_a$ & EP-B \cite{garcia_duran_learning_2017} & 128 & 1 & 78.1 & \underline{71.0} & \textbf{80.0} \\
\hline
\textbf{\textsc{RAE} (this paper)} & & & & & \\
$A, X_a$      & Untrained    & 256 & 2 &      $57.0 \pm 2.7\%$   &     $46.0 \pm 3.4\%$      &   $ 65.7 \pm 2.3\%$                              \\
$A, X_a$      & $J_{within}$ (only shuffling)  & 256 & 2 & $77.5 \pm 1.1\%$  & $67.2 \pm 1.0\%$                & $76.9 \pm 1.8\%$ \\
$A, X_a$      & $J_{within}$  & 256 & 2 & $78.0 \pm 0.9\%$     & $67.6 \pm 1.1\%$                & $77.0 \pm 1.8\%$ \\
$A, X_a$      & $J_{neighbor}$ & 256 & 2 & $79.3 \pm 1.2\%$     & $68.1 \pm 1.3\%$                & $\underline{79.5} \pm 0.7\%$ \\
$A, X_a$      & \textsc{concat}($J_{within}, J_{neighbor}$)        & 256 + 256 & 2 & $
\underline{80.2} \pm 0.9\%$ & $68.0 \pm 1.2\%$           & $78.7 \pm 0.9\%$ \\
\hline
\textbf{Semi-supervised models} & & & & & \\
$A, X_a, Y$      & Planetoid \cite{yang_revisiting_2016} & -  &  - & 75.7            & 64.7               & 77.2                          \\
$A, X_a, X_s, Y$ & GraphSAGE  \cite{chen_stochastic_2017} & 32 & 2 & 82.0         & 70.8                          & 79.0                    \\
$A, X_a, Y$   & GCN \cite{kipf_semi-supervised_2016}               & 16 & 2 & 81.5               & 70.3                          & 79.0                          \\
$A, X_a, Y$   & GAT  \cite{velickovic_graph_2018}               & 64 & 3 & 83.0   & 72.5                 & 79.0    \\
$A, X_a, Y$   & FastGCN-transductive  \cite{chen_fastgcn:_2018}               & 16  & 2 & 81.8   &         -         & 77.6    \\
\hline
\end{tabular}
\end{center}
\end{adjustwidth}
\caption{Accuracy and mean standard errors of various methods on three standard node classification tasks. The left column indicates which information is available to the algorithm: $A$ the adjacency matrix, $X_s$ structural features, $X_a$ action features, $Y$ the outcome of interest. $d$ is either the embedding dimension in the unsupervised case, or the dimension of the last hidden layer in the supervised case. $K$ is the maximum depth in the network that the model has access to. The first block presents unsupervised baselines, the second block presents results from our modeling architecture, the third block presents semi-supervised models. The best \emph{unsupervised} model is bolded, and the second best is underlined.}
\label{table:node_results}

\end{table*}

\begin{figure*}
 \centering
 \includegraphics[width=\textwidth]{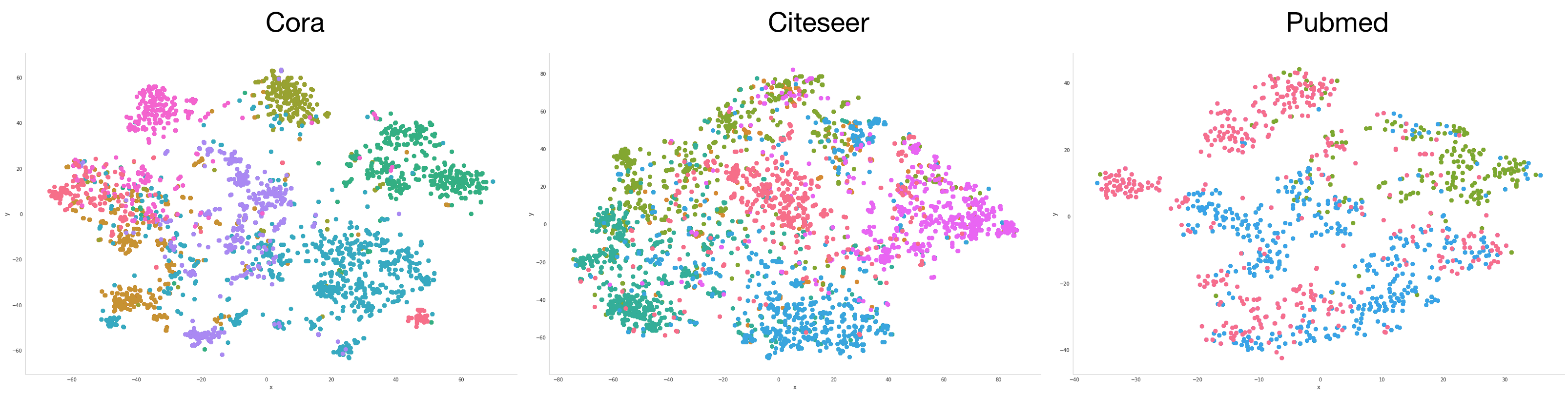}
 \caption{Visualization of embeddings on three citation graphs using \textsc{RAE} with  $J_{within}$, reduced to two dimensions using TSNE. For Cora and Citeseer, we used the entire dataset, while we visualize the test set for Pubmed. Clusters are quite distinct in all cases. On Cora, this model achieves a silhouette score of 0.166.}
 \label{figure:viz}
\end{figure*}

We describe model architecture in Algorithms \ref{algorithm:RAE} and \ref{algorithm:meanagg}. We choose depth $K = 2$, and sample sample 10 neighbors at depth $K=1$ and 25 neighbors at depth $K=2$, for a total receptive field of 261 nodes (ego + 10 at depth 1 + 250 at depth 2). We use a dropout rate of 0.6, a constant learning rate of 0.01, and a batch size of 256. Unless noted, models are trained for 200 epochs, where each epoch considers a single batch. This means that even though graph sizes vary, a constant 51,200 training examples are used in all cases. We choose an output layer dimension of $d_{K=1}=64$ and a hidden layer representation of $d_{K=2} = 192$. The embedding dimension is therefore $d = 256$, since it's a concatenation of the representation from $K=1$ and $K=2$.

The model is trained with 5 positive examples and 20 negative examples, using $J_{within}$ or $J_{neighbor}$ directly with no margin. When using $J_{within}$, half of the negative examples are generated via neighbor shuffling and half are generated randomly from $R(u)$. For $J_{neighbor}$, all negative examples are generated from non-neighbors with $P(u)$.  Models are trained with Adam \cite{kingma_adam:_2014} and initialized with Xavier uniform weights \cite{glorot_understanding_2010}.

In the case of sparse input vectors representing text in citation networks, no preprocessing is performed. This means directly inputting high-dimensional (500 to 3703) sparse vectors into the model. Following common practice, we use embeddings in an L2-penalized logistic regression model. In the multiclass case, a one-vs-rest classifier is used. To obtain an accurate picture of both performance and variance, we report results from 20 runs, providing mean performance and standard deviation of the mean\footnote{An estimate of the standard error of the mean can be obtained by dividing by the square root of the number of runs.}.

Our choices of $d=256$, $K=2$ and a receptive field of 261 are equivalent or conservative compared to prior research on GNNs. For instance, \cite{velickovic_deep_2018} use $d=512$ and $K=3$, and \cite{hamilton_inductive_2017} use $d=256$ and $K=2$. Since the output dimension for \textsc{RAE} is set to $d_{K=1} = 64$, we use many fewer parameters than a \textsc{GraphSAGE}-style model which has both hidden and output dimensions of 256.

\subsection{Graph classification results}

We choose six standard datasets studied in the graph classification literature (full dataset descriptions can be found in \cite{yanardag_deep_2015}), with results presented in Table \ref{table:graph_results}. Most of these datasets do not have features beyond those from the graph itself\footnote{The only exception is MUTAG, where nodes are in one of 7 categories. In this case we create a one-hot encoding of node category and append it to the feature vector generated from node degrees.}. In these cases, we label each node with the sorted degrees of its neighbors, capped at 30. When a node has more than 30 neighbors, we sample randomly; when it has fewer, we pad with zeros. For MUTAG, IMDB-B, and REDDIT-B only sample 4 neighbors are sampled at depth 1 and 4 at depth 2, for a total receptive field of 21. For the rest of the graphs we use the parameters described above. We observed that training did not substantially improve performance, so these models are untrained. They therefore indicate the raw expressive power of the model. A discussion of different GNN architectures and theoretical performance guarantees can be found in \cite{xu_how_2018}. 

Since each node has a $d$-dimensional vector, we need to choose a ``readout'' function to produce graph vectors from node vectors. We choose a simple summation to represent the embedding for graph $G_i$: $z_{G_i} = \sum_{u \in G_i} z_u$. Graph vectors $z_{G_i}$ are then employed in a one-vs-rest classification problem using 10-fold cross validation. In the binary classification case, we use the test folds to select the probability cutoff which maximizes test fold accuracy.

\textsc{RAE} outperforms two strong baselines on many of the graph classification tasks: deep graph kernels \cite{yanardag_deep_2015} and \textsc{Patchy-san} \cite{niepert_learning_2016}. We also include graph isomorphism networks (GIN) \cite{xu_how_2018} to give an idea of the upper bound for model expressiveness if we were to use one-hot encodings for features, a summation aggregator, training, and a larger $K$. The simple mean framework we use performs quite well.

\subsection{Node classification results}

We choose three standard datasets which provide both graphs and rich features: Cora, Citeseer, and Pubmed. Dataset details can be found in Table \ref{table:node_datasets}. These are citation datasets where each node is a paper and each link is a citation. Each paper is annotated with a sparse vector representing the words used in the paper. We follow the procedure described in \cite{kipf_semi-supervised_2016} for these datasets: 20 nodes are selected as training examples from each class, with 1000 nodes selected as test nodes and 500 as validation. We use the same splits used by Kipf and Welling\footnote{The specific splits can be found here \url{https://github.com/tkipf/gcn}.}. We optimized hyperparameters on the validation set given by Kipf and Welling, but do not make use of validation set on the fly to determine early stopping. Hyperparameters were tuned on the Cora dataset and then applied these settings directly to Citeseer and Pubmed.

Table \ref{table:node_results} presents results of \textsc{RAE} compared to a variety of other unsupervised and semi-supervised baselines. The most direct comparison is with other unsupervised methods. All versions of \textsc{RAE} outperform DeepWalk-based methods (including node features directly). Moving to two more difficult baselines, \textsc{RAE} outperforms Deep Graph Infomax (DGI) \cite{velickovic_deep_2018} on Pubmed and EP-B \cite{garcia_duran_learning_2017} on Cora. DGI has access to depth 3 rather than the depth 2 used in our model, making the performance of \textsc{RAE} impressive. EP-B is an efficient method which does a single aggregation step, but also comes with a margin hyperparameter which must be tuned, and samples all neighbors of a focal node as opposed to the fixed number used here.

Surprisingly, \textsc{RAE} with $J_{neighbor}$ outperforms the supervised models on Pubmed, and scores the second highest overall. These results indicate that the modeling framework proposed here provides strong performance on these standard tasks despite model simplicity. $J_{within}$ using only neighbor shuffling---the most scalable parameterization of \textsc{RAE}---performs comparably to DGI on Pubmed and only a shade below EP-B on Cora. The most puzzling part of these results is the weak performance on Citeseer, which could be related to the larger input dimension (3703). Regularization may help in this setting, but was not employed since it did not prove effective when developing the model on the Cora validation set.

The embeddings presented here do not incorporate structural features (e.g. degree, neighbor degrees, motif counts) in the input vector \cite{hamilton_representation_2017,henderson_rolx:_2012,ahmed_learning_2018}. We therefore refer to them as \textit{action} embeddings. We consistently found that including even node degree in the feature vector reduced performance. This implies that unsupervised GNNs are not adept at preserving both structural and action features without additional modeling work. Structural features easily overwhelm sparse action features.

The combination of $J_{within}$ and $J_{between}$ provides the strongest performance on Cora, but not on the other two datasets. While this may seem surprising, it is potentially related to the small size of the training set (20 nodes per class). As a final note, we have considered only transductive embeddings here\footnote{This is both to compare more closely with previous work and due to space.}, but \textsc{RAE} can be applied inductively as well.

\section{Conclusion}

We have presented \textsc{RAE}, a scalable methodology for learning role embeddings, and introduced the concept of \textit{role action embeddings}. This creates a useful distinction in the type of information which may be incorporated into node embeddings, in addition to providing strong performance on several standard tasks.

We note that the performance of action feature embeddings is related to the type of task under consideration: when classifying papers into topic categories, it makes sense that paper text is important. However, without knowledge of how embeddings will be used beforehand, this distinction creates a challenge for researchers. A simple response is to train separate models for structural and action features. A direction for future work is to combine these two types of features with novel model architectures. In addition to the intuitive usefulness of role action embeddings, the modeling choices we have made are likely to be useful in other settings as well.

\section{Acknowledgements}

Thanks to Nima Noorshams, Katerina Marazopoulou, Lada Adamic, Saurabh Verma, Alex Dow, Shaili Jain, and Alex Pesyakhovich for discussions and suggestions.

\bibliographystyle{ACM-Reference-Format}
\bibliography{rolesage}

\end{document}